\documentstyle[epsf]{mn}

\newif\ifAMStwofonts

\def\simlt{\lower.5ex\hbox{$\; \buildrel < \over \sim \;$}}
\def\simgt{\lower.5ex\hbox{$\; \buildrel > \over \sim \;$}}
\def\boom{BOOMERanG }

\ifoldfss
  \ifCUPmtlplainloaded \else
    \NewTextAlphabet{textbfit} {cmbxti10} {}
    \NewTextAlphabet{textbfss} {cmssbx10} {}
    \NewMathAlphabet{mathbfit} {cmbxti10} {} 
    \NewMathAlphabet{mathbfss} {cmssbx10} {} 
  \fi
  \ifAMStwofonts
    \ifCUPmtlplainloaded \else
      \NewSymbolFont{upmath} {eurm10}
      \NewSymbolFont{AMSa} {msam10}
      \NewMathSymbol{\upi}     {0}{upmath}{19}
      \NewMathSymbol{\umu}     {0}{upmath}{16}
      \NewMathSymbol{\upartial}{0}{upmath}{40}
      \NewMathSymbol{\leqslant}{3}{AMSa}{36}
      \NewMathSymbol{\geqslant}{3}{AMSa}{3E}

      \let\leq=\leqslant \let\le=\leqslant
       
    \fi
  \fi
\fi 

\ifnfssone
  \newmathalphabet{\mathit}
  \addtoversion{normal}{\mathit}{cmr}{m}{it}
  \addtoversion{bold}{\mathit}{cmr}{bx}{it}
  \newmathalphabet{\mathbfit} 
  \addtoversion{normal}{\mathbfit}{cmr}{bx}{it}
  \addtoversion{bold}{\mathbfit}{cmr}{bx}{it}
  \newmathalphabet{\mathbfss} 
  \addtoversion{normal}{\mathbfss}{cmss}{bx}{n}
  \addtoversion{bold}{\mathbfss}{cmss}{bx}{n}
  \ifAMStwofonts
    \ifCUPmtlplainloaded \else
      \UseAMStwoboldmath
      \makeatletter
      \new@mathgroup\upmath@group
      \define@mathgroup\mv@normal\upmath@group{eur}{m}{n}
      \define@mathgroup\mv@bold\upmath@group{eur}{b}{n}
      \edef\UPM{\hexnumber\upmath@group}
      \new@mathgroup\amsa@group
      \define@mathgroup\mv@normal\amsa@group{msa}{m}{n}
      \define@mathgroup\mv@bold\amsa@group{msa}{m}{n}
      \edef\AMSa{\hexnumber\amsa@group}
      \makeatother
      \mathchardef\upi="0\UPM19
      \mathchardef\umu="0\UPM16
      \mathchardef\upartial="0\UPM40
      \mathchardef\leqslant="3\AMSa36
      \mathchardef\geqslant="3\AMSa3E

      \let\leq=\leqslant \let\le=\leqslant

    \fi
  \fi
\fi 

\ifnfsstwo
  \DeclareMathAlphabet{\mathbfit}{OT1}{cmr}{bx}{it}
  \SetMathAlphabet\mathbfit{bold}{OT1}{cmr}{bx}{it}
  \DeclareMathAlphabet{\mathbfss}{OT1}{cmss}{bx}{n}
  \SetMathAlphabet\mathbfss{bold}{OT1}{cmss}{bx}{n}
  \ifAMStwofonts
    \ifCUPmtlplainloaded \else
      \DeclareSymbolFont{UPM}{U}{eur}{m}{n}
      \SetSymbolFont{UPM}{bold}{U}{eur}{b}{n}
      \DeclareSymbolFont{AMSa}{U}{msa}{m}{n}
      \DeclareMathSymbol{\upi}{0}{UPM}{"19}
      \DeclareMathSymbol{\umu}{0}{UPM}{"16}
      \DeclareMathSymbol{\upartial}{0}{UPM}{"40}
      \DeclareMathSymbol{\leqslant}{3}{AMSa}{"36}
      \DeclareMathSymbol{\geqslant}{3}{AMSa}{"3E}

      \let\leq=\leqslant \let\le=\leqslant

    \fi
  \fi
\fi 

\ifCUPmtlplainloaded \else
  \ifAMStwofonts \else 
    \def\upi{\pi}
    \def\umu{\mu}
    \def\upartial{\partial}
  \fi
\fi

\title[How old is the Universe~?]
	{How old is the Universe~?\\ Setting new constraints 
	on the age of the Universe}
\author[Ferreras, Melchiorri \& Silk]
{Ignacio Ferreras, Alessandro Melchiorri and Joseph Silk\\
Nuclear \& Astrophysics Lab. 1 Keble Road, Oxford OX1 3RH, United Kingdom}
\date{Draft version \today}
\pagerange{\pageref{firstpage}--\pageref{lastpage}}
\pubyear{2001}

\begin{document}

\maketitle

\label{firstpage}

\begin{abstract}
There are three independent techniques for determining the age of
the universe: via cosmochronology  of long-lived radioactive nuclei,
via stellar modelling and population synthesis of the oldest
stellar populations, and,
most recently, via  the  precision cosmology that has become
feasible with the mapping of the acoustic peaks in the
cosmic microwave background. 
We demonstrate that all three methods give completely consistent results,
and enable us to set rigorous bounds on the maximum and minimum ages  
that are allowed for the universe.
We present new constraints on the age of the universe 
by performing a multiband colour analysis of bright cluster ellipticals over a large
redshift range ($0.3<z<0.9$), which allows us to infer the ages of 
their stellar populations
over a wide range of possible formation redshifts and metallicities. 
Applying a conservative prior to Hubble's
constant of $H_0=70\pm 15$ km s$^{-1}$ Mpc$^{-1}$, we find the age
of the universe to be $13.2^{+3.6}_{-2.0}$~Gyr ($1\sigma$), in 
agreement both with the estimates from type~Ia supernovae,
as well as with the latest uranium decay estimates, which yield  an
age for the Milky Way of $12.5\pm 3$~Gyr. If we combine the results
from cluster ellipticals with the analysis of the angular power
spectrum of the cosmic microwave background and with the observations of
type~Ia supernovae at high redshift, we find a similar age:
$13.4^{+1.4}_{-1.0}$~Gyr. Without the assumption of any
priors, universes older than $18$~Gyr are ruled out by the data 
at the 90\% confidence level.
\end{abstract}



\section{Introduction}
Discrepancies between  age determinations, such as for globular clusters
and from the Hubble constant, have long plagued cosmology. The situation 
has changed dramatically in  the past three years, however, as the 
Hubble constant proponents (Mould et al. 2000) have converged on a 
value which we
conservatively take to be of $H_0=70\pm 15$ km s$^{-1}$ Mpc$^{-1}$, 
and because of several  other developments. Type~Ia supernova
measurements  have provided strong evidence for a cosmological
constant (Perlmutter et al. 1999; Riess et al. 1998), 
and mapping of the acoustic peaks in
the cosmic microwave background has fixed the cosmological parameters with
unprecedented precision (Netterfield et al. 2001). 
Finally, detection of uranium in  
an old Population~II star has provided a  direct nuclear chronometer
for the age of our galaxy (Cayrel et al. 2001). 
Any one of these measurements 
may be suspect, but the remarkable concordance that we find enables 
to show here by a combined likelihood anaysis that combination of 
these constraints provides, for the first time, rigorous upper and 
lower bounds on the age of the universe. 

Consider first stellar model determinations of the age of the universe.
The small scatter found in the absolute luminosity of the brightest
cluster galaxies (BCGs) over a wide range of redshifts motivated their
use as standard candles to determine cosmological parameters 
(Gunn \& Oke 1975). However, this analysis was based on the
assumption that this type of galaxy should not undergo significant
evolution in luminosity with lookback time. Tinsley (1976)
showed that the brightening of main sequence stars poses a
major hurdle in the use of BCGs as standard candles.
Predicting the luminosity evolution is a rather challenging endeavour 
since it strongly depends both on the star formation history as 
well as on the dynamical history. An analysis of bright
ellipticals in a large sample of clusters observed in the
near-infrared (Arag\'on-Salamanca, Baugh \& Kauffmann 1998)
concluded that the stellar mass in BCGs over a large redshift 
range ($0<z<1$) has evolved by a factor between $2$ and $4$ 
depending on the cosmology, in agreement with the predictions 
of hierarchical clustering scenarios of structure 
formation (Kauffmann \& Charlot 1998). On the other hand, an 
analysis of the colours of bright cluster ellipticals is only 
dependent on their star formation history. The small scatter found 
in the colour-magnitude relation of cluster ellipticals
(Bower, Lucey \& Ellis 1992; Stanford, Eisenhardt \& Dickinson 1998)
hints at old stellar populations, formed at redshifts $z_F\simgt 3$.
The use of galaxy colours as a ``cosmic clock'' is nevertheless
a challenging task especially due to the age-metallicity degeneracy,
which causes age effects to be mimicked by a range of metal 
abundances (Worthey 1994).

We focus on the brightest cluster ellipticals
--- which are assumed to have a simple star formation history as
explained below --- and use stellar population synthesis models 
(Bruzual \& Charlot, in preparation, hereafter B\&C) 
in order to infer the age of the
stars in these galaxies. 
We have compared this technique both with the analysis of the 
latest measurements of the angular power spectrum 
of the cosmic microwave background radiation (CMBR)  observed by the 
\boom collaboration (de~Bernardis et al. 2000, 2001) as well as with the 
observations of high redshift type~Ia supernovae
(Perlmutter et al. 1999; Riess et al. 1998)
in order to estimate the age of the universe. We also compare these 
ages with the ages derived for globular clusters (Salaris \& Weiss 1998)
and the oldest halo stars (Cayrel et al. 2001), to which 
one has to add an age for the Milky Way that must correspond 
to the time elapsed between the Big Bang and formation 
at a redshift of at least $2$, and more conservatively
$5$ or even $10$.

We find that  four completely independent age determinations ---
namely  CMBR, galaxy colours for clusters at $z\simlt 1$, stellar 
evolution applied to old globular clusters, and radioactive isotope 
dating of old stars --- lead to a consistent result.
These age probes represent 
different combinations of Big Bang parameters
($H_0$, $\Omega_m$, $\Omega_\Lambda$), galaxy evolution parameters
(star formation rate history, initial stellar mass function),
stellar evolution parameters (stellar mass, composition
and mixing length), and nucleochronology (half-life of $^{238}U$),
respectively.  In this paper we add the spectrophotometric
study of bright cluster ellipticals to the growing list of
cosmological probes used to determine the 
age of the Universe (Lineweaver 1999; Primack 2000).

\section{Using bright cluster ellipticals as cosmic clocks}
We use the sample of Stanford, Eisenhardt \& Dickinson (1998),
which comprises~17 clusters over a large redshift range ($0.3<z<0.9$).
The sample was extracted on the basis of available imaging with 
the Wide Field and Planetary Camera~2 on board the Hubble 
Space Telescope
--- for morphological classification purposes --- from a larger
sample of 46 clusters drawn from a variety of optical, X-ray and
radio-selected clusters. Each cluster was imaged in near-infrared
$J$, $H$ and $K$ bands
as well as two optical passbands, which were chosen as a function
of redshift to straddle the 4000\AA\  break in the galaxy rest-frame
(i.e. roughly mapping rest-frame $U$ and $V$ bands).
For each cluster we select the three brightest early-type systems
which fall on the colour-magnitude relation and choose the reddest 
one, always taking care not to select an outlier. For each galaxy we 
compared the observed colours with the predictions of a grid
of simple stellar populations with different ages and metallicities,
from the latest models of B\&C. The comparison
was performed using a $\chi^2$ test applied between the four observed 
colours for each of the~17 clusters ($c_{n,i}$; $n=\{1\cdots 17\}$, 
$i=\{1\cdots 4\}$), and the predictions from the
population synthesis models (${\cal C}_{n,i}$), namely:
\begin{equation}
 \begin{array}{rcl}
\chi^2_{SSP} & = & \sum_{n=1}^{17} \sum_{i=1}^4 \big( 
c_{n,i}-{\cal C}_{n,i}\big)^2/\sigma_i^2 +\\
 & & + \big({\rm Model}(U-V)_{z=0} - 1.6\big)^2/\sigma_0^2.
\end{array}
\end{equation}
The colour scatter $\sigma_i$ is chosen to be $\pm 0.2$ magnitudes for all
four colours, and comprises the effect of photometric error bars, 
uncertainties in the modelling of stellar populations as well as 
colour scatter in bright ellipticals. In order to further tighten the 
allowed region of parameter space, we have applied a constraint at
zero redshift, using high precision photometry of the Coma and
Virgo clusters (Bower et al. 1992), adding the second term
shown in equation (1). Hence, for a given age-metallicity point, 
we evolve the system to
zero redshift (with a given cosmology) and then compare its
$U-V$ colour with the observed $(U-V)_{z=0}=1.6$ with a scatter
$\sigma_0=0.1$~mag.

Figure~1 shows the contours at 68, 90 and 95\% confidence levels in
an age-metallicity diagram for the brightest ellipticals in cluster 
Cl0024+16 ($z=0.39$). The degeneracy between age and metallicity is
readily shown, so that old populations with low metal abundances
give similar results to young stars with a higher metallicity. In 
the figure, we compare the result of B\&C with the population synthesis
models of Yi et al. (1999) who take special care in adding the
contribution from core helium burning stars --- 
horizontal branch (HB) stars --- and their progeny. 
HB stars seem to be the most plausible candidate to explain 
the UV upturn ($\lambda\sim 1500$\AA ) in elliptical 
galaxies (O'Connell 1999). However, the agreement between both models
shown in Figure~1 is expected if we consider that most of 
the contribution of the light in the spectral range considered
(i.e. between $U$ and $K$ bands) from elliptical galaxies comes
from the main sequence and red giant branch populations, which are
better understood than HB stars (Bruzual \& Charlot 1993;
Yi et al. 1999).

We can impose a further constraint on the metallicity for bright
cluster ellipticals. In these massive systems, the deep gravitational
wells prevent gas ejected from supernovae from being thrown out
of the galaxy. This feedback mechanism is presumed to be more effective
in low mass spheroids (Ferreras \& Silk 2000a)  and is the basis for the
correlation between mass and metallicity (Larson 1974). A detailed 
analysis of metal abundances in several Fornax cluster early-type
galaxies (Kuntschner 2000) shows that $[{\rm Fe/H}]\simgt 0.0$ for 
ellipticals with high velocity dispersion ($\log\sigma\simgt 2.2$).
The standard simple closed-box model is a good approximation to 
the chemical enrichment in bright ellipticals (Pagel 1997).
In this model the evolution of the average stellar metallicity is:
\begin{equation}
Z = p \Big[ 1 + \frac{\mu\ln\mu}{1-\mu}\Big],
\end{equation}
where $\mu$ is the gas mass fraction contributing to star formation 
and $p$ is the stellar yield, i.e. the mass fraction of elements 
other than helium generated in stars and weighted by the initial 
mass function (IMF). In bright ellipticals there is no significant ongoing 
star formation, which means $\mu\rightarrow 0$ and so
$Z\sim p\simlt 0.8 Z_\odot$ (where $Z_\odot = 0.02$ is the solar
metallicity). Hence, we can further constrain the upper 
bound to the age of the stellar populations by imposing a lower 
limit to the metallicity at $[{\rm Fe/H}]\simgt -0.1$, roughly corresponding 
to $Z\sim 0.8Z_\odot$, shown as a horizontal line in Figure~1.

Finally, the complete sample of $17$ clusters can be combined
by assuming a given cosmology $(H_0, \Omega_m,\Omega_\Lambda )$ 
which enables us to translate redshifts into ages.
A grid of models was run as follows:
The cosmology was explored by choosing $0.085\leq\Omega_m\leq 1.085$
and $0\leq\Omega_\Lambda\leq 1.$, both in steps of $0.085$; and
$0.25\leq h_0\leq 0.95$ in steps of $0.05$ (with $h_0$ defined as
$H_0/$km s$^{-1}$Mpc$^{-1}$). The stellar populations were parametrized
by the metallicity, chosen in the range $0.8\leq Z/Z_\odot\leq 1.9$
in steps of $0.1$ and the formation redshift $z_F=\{2, 2.25, 2.50, 2.75, 
3, 3.25, 3.50, 3.75, 4, 6, 8, 10\}$. We ran two such models for two
different initial mass functions: Salpeter (1955) and Scalo (1986),
using B\&C and a third one using the models of Yi et al. (1999)
for a Salpeter IMF.

The metallicity is not the only factor that could modify age estimates
from broadband photometry. The reddening caused by dust may also
cause overestimates of stellar ages. 
However, dust is not expected
to play a significant role in cluster ellipticals, whose gaseous
component is too hot to allow significant amounts of dust to be present
over large timescales. The small scatter found in the colour-magnitude 
relation of cluster ellipticals would require a tight conspiracy
between age, metallicity and dust, in order to keep the scatter 
as low as observed. Furthermore, rest-frame near-ultraviolet photometry
of Abell~851 ($z=0.41$) shows that the brightest ellipticals are not
redder than the dustless predictions of simple stellar 
populations (Ferreras \& Silk 2000b).

\section{Age from cosmology: method}
In the {\it standard} inflationary framework, 
the cosmic microwave background power spectrum depends essentially
on 3 cosmological parameters (Efstathiou \& Bond 1999): 
the physical matter density
in baryons ($\omega_b=\Omega_bh^2$), the overall physical matter 
density ($\omega_{m}=\Omega_{m}h^2$) and the parameter 
$R =\sqrt{\omega_m / \omega_k} f(y)$ where $f(y)$ is $\sinh (y)$,
$y$, $\sin (y)$ for open, flat and closed models respectively and 
where
\begin{equation}
y=\omega_k^{1/2}\int_{a_r}^1{{da \over 
{[\omega_ma+\omega_ka^2+\omega_{\Lambda}a^4]^{1/2}}}},
\end{equation}
with $\omega_k=h^2(1-\Omega_m-\Omega_{\Lambda})$.
Cosmological models with the same values of the parameters
$\omega_m$, $\omega_b$ and $R$ will have nearly-identical power
spectra on degree and subdegree angular scales.
Furthermore, the age of the universe is given by
\begin{equation}
t_0=9.8 {\rm Gyr}\int_0^1{{a da \over 
{[\omega_ma+\omega_ka^2+\omega_{\Lambda}a^4]^{1/2}}}}.
\end{equation}

Under the assumption of a flat universe 
(i.e. $\omega_k = 0$ and $R \sim const$) as the
recent CMBR measurements seem to suggest, it is easy to show
that nearly degenerate models with the same $\omega_{\Lambda}$, 
$\omega_m$ and $\omega_b$ will have also similar ages. 
Thus, in principle, a measurement of the CMBR spectrum can be 
extremely helpful in the determination of the age of the universe
in flat cosmologies.
The curvature itself, however, due to the dependence of $R$ from
$\Omega_{\Lambda}$ and $\Omega_m$, cannot be constrained by CMBR
measurements alone better than $10-20 \%$.

This introduces a limitation in the use of the CMBR spectrum
for producing independent strong constraints on the age.
However, the integrated Sachs-Wolfe effect on large angular scales
and the assumption of mild external priors on the various
cosmological parameters can break the above degeneracies and
reduce the error in the age estimation from CMBR.
In what follows we will put constraints on the age of the universe
by comparing the recent CMBR data obtained
from the BOOMERanG experiment with a database of models with 
cosmological parameters sampled as described in the previous section.
We also vary the spectral index of the primordial density perturbations
within the range $n_s=0.50, ..., 1.50$, the optical depth 
$\tau_c= 0.0, ..., 0.3$, and we rescale the fluctuation amplitude by a
pre-factor taken as a free parameter.

The theoretical models are computed using the publicly available 
{\sc cmbfast} program (Seljak \& Zaldarriaga 1996) and are compared with the
BOOMERanG-98 and COBE results. We include the COBE data using Lloyd 
Knox's RADPack packages.
The power spectra from these experiments are estimated in
$19$ and $24$ bins respectively, spanning the range
$2 \le \ell \le 1050$. For the BOOMERanG-98 the spectrum we 
assign a flat shape, $\ell(\ell+1)C_{\ell}/2\pi=C_B$.

Following de~Bernardis et al. we approximate 
the signal $C_B$ inside the bin as a Gaussian variable. 
The likelihood for a given cosmological model is then
defined by $L=e^{-\chi^2_{CMBR}/2}$ with
\begin{equation}
\chi^2_{CMBR} =\sum_{B} (C_B^{th}-C_B^{ex})^2/\sigma_B^2,
\end{equation}
where $C_B^{th}$ ($C_B^{ex}$) is the theoretical (experimental)
band power, and $\sigma_{B}$ is the quoted error bar. 
We consider a $10 \%$  calibration error for the 
BOOMERanG-98 experiment by adding a gaussian term 
$\chi^2_{cal} = (1.0-A_{cal})^2/(0.24)^2$ 
and by finding the value of  $A_{cal}$ that for
a given cosmological models maximizes the likelihood.
We also marginalize over the beam uncertainty ($1.4^\prime$) and we
found that the removal of the last 3 bins 
--- which are more likely to be affected by systematics ---
does not change the results of our analysis.
We multiply $L$ by our chosen priors and 
attribute a likelihood to each age in the $1-30$~Gyr range
by finding the 'nuisance' parameters that
maximise it. We then define our central values and $1\sigma$
for the age from the $16 \%$, $50 \%$ and $84 \%$ integrals
of $L$ over age.

\section{Discussion}
The results are summarized in Table~1 using external priors 
based on theoretical restrictions as well as on recent 
astronomical observations. The finite volume of parameter space 
sampled imposes further implicit constraints on the age of
the universe.
However, the large range of parameters explored
implies this implicit constraint is rather weak, as can be
seen from the last entry in Table~1 (constant likelihood
across the parameter space without any priors), for which the estimated
age of the universe at a 68\% confidence level is $19.8\pm 9$~Gyr.
The constraint imposed by stellar populations in bright cluster
ellipticals (Salpeter IMF) yields an age between $12$ 
and $13$~Gyr regardless of the prior on $h_0$ or on the 
population synthesis model chosen, 
although imposing a prior on $h_0$ results in smaller error bars, 
roughly around $13.2^{+3.6}_{-2.0}$~Gyr. This is in agreement with estimates 
of the age of the universe from Type~Ia supernovae at high 
redshift (Perlmutter et al. 1999; Riess et al. 1998). Using a Scalo
IMF does not change the age estimates significantly: the best fit
gives ages less than $5\%$ compared to a Salpeter IMF, i.e. well 
below the error bars.

The latest measurement of the age of the oldest
stars in the Milky Way through the decay of $^{238}$U gives a value
of $12.5\pm 3$~Gyr (Cayrel et al. 2001), which is consistent with 
the above ages if we assume the process of star formation started 
in our galaxy $1-3$~Gyrs after the Big Bang, corresponding to a 
formation redshift $z_F\simgt 2$ for a reasonable range of cosmologies. 
Another technique which allows for a reasonably accurate estimate of the
age of our galaxy involves globular clusters.
The ages of globular clusters can be inferred in a distance-independent
way by analyzing several features in the stellar colour-magnitude diagram
such as the luminosity gap between the main sequence turnoff and the
base of the zero age horizontal branch, or the colour gap between
the turnoff point and the tip of the red giant branch. Salaris 
\& Weiss (1998) explored a sample of halo and disk globular clusters,
finding the age of the oldest cluster (NGC~6366) to be 
$12.2\pm 1.1$~Gyr. However, Pont et al. (1998) estimate an
age of $14$~Gyr for M92 with a more standard distance-dependent
technique, involving a fit to the main-sequence in a colour-magnitude
diagram. M92 is estimated by Salaris \& Weiss (1998) to be 
$11.0\pm 1.1$~Gyr old. 

The estimates from CMBR data give values that are perfectly consistent 
with the analysis using stellar populations. 
Since the new BOOMERanG data is perfectly consistent with present
estimates of the baryon content from Big Bang Nucleosynthesis 
(Burles et al. 1999) the prior $\Omega_bh^2 = 0.02\pm 0.002$ has little 
effect on the results as we can see in the fifth row of Table~1.
We note that our CMBR analysis is restricted
to a specific class of models based on {\it adiabatic} 
primordial perturbations and with a limited number of parameters. 
Assuming different mechanisms of structure formation than
those predicted by inflation such as topological defects and/or
isocurvature fluctuations would drastically change
our conclusions. Furthermore, restricting our analysis to
purely baryonic universes would yield higher values for the 
age of the universe, namely around $\sim 22$~Gyr 
(Griffiths, Melchiorri \& Silk 2001).
Thus, the consistency between the age values inferred from CMBR and
those obtained by stellar populations can be considered as a
further confirmation of the standard inflationary scenario.

As an interesting check of the models, we decided to invert the 
analysis presented here so that the formation
redshift and the average metallicity of stellar populations in 
bright ellipticals could be inferred from estimates to the age
of the universe. With a conservative prior on Hubble's constant 
($H_0=70\pm 15$~km~s$^{-1}$~Mpc$^{-1}$), we find the
average metallicity to be $Z/Z_\odot = 1.10\pm 0.18$ and 
$1.17\pm 0.31$ for a Salpeter and a Scalo IMF, respectively,
using the population synthesis models of Bruzual \& Charlot. 
Adding the age constraint from CMBR data 
does not significantly change 
the result, giving $Z/Z_\odot=1.14\pm 0.18$ (Salpeter)
and $1.14\pm 0.28$ (Scalo). On the other hand, all formation redshifts 
explored in this paper ($2< z_F< 10$) are allowed, and 
only the latest formation epochs ($z_F\sim 2$) are mildly ruled
out at the $68\%$ confidence level, which is compatible with the
latest estimates from morphological and spectroscopic studies of
high redshift clusters (Van~Dokkum \& Franx 2001).

Hence, we have presented the study of the
colours of bright cluster ellipticals as an additional
analysis to be incorporated in the medley of cosmological 
probes. With a set of reasonable assumptions for the
stellar populations in this type of galaxies, we infer 
an age of the universe of $13.2^{+3.6}_{-2.0}$~Gyr and a final result
--- combining the results from stellar populations in cluster 
ellipticals, the angular power spectrum of the CMBR and type~Ia 
supernovae --- of $13.4^{+1.4}_{-1.0}$~Gyr. Without the 
assumption of any priors, the combined analysis rules out 
universes older than $18$~Gyr at a 90\% confidence level.

\section*{Acknowledgments}
IF is supported by a grant from the European Community under 
contract HPMF-CT-1999-00109.

\newpage
\onecolumn
\begin{table}
{\bf TABLE 1: Constraints on the Age of the Universe}
\begin{center}
\begin{tabular}{ccccccc}
Prior & B\&C & Yi & CMBR & CMBR+B\&C & CMBR+B\&C+SN-Ia & Database\\\hline

$h=0.70 \pm 0.15$,$\omega_b=0.025\pm0.01$ & 
$13.2^{+3.6}_{-2.0}$ & $13.4_{-2.8}^{+4.4}$ &
$13.8^{+1.8}_{-1.4}$ & $13.6^{+1.6}_{-1.0}$ & $13.4^{+1.4}_{-1.0}$ &
$16.0_{-6.0}^{+6.4}$\\

$h=0.72 \pm 0.08$,$\omega_b=0.025\pm0.01$ & 
$13.0^{+3.0}_{-2.0}$ & $13.2_{-2.4}^{+4.4}$ & 
$13.2^{+1.2}_{-1.0}$ & $13.2^{+1.0}_{-0.8}$ & $13.2^{+1.2}_{-0.8}$ &
$14.2_{-4.4}^{+4.4}$\\

$h=0.70 \pm 0.15$, $\omega_b=0.025\pm0.01$, $\Omega=1$ & 
$13.2^{+3.4}_{-2.2}$ & $13.6_{-3.0}^{+4.0}$ &
$13.4^{+1.2}_{-1.0}$ & $13.4^{+1.0}_{-0.8}$ & $13.4^{+1.0}_{-1.0}$ &
$15.2_{-5.4}^{+6.8}$\\

$h=0.70 \pm 0.15$,$\omega_b=0.02\pm0.002$ & 
$13.0^{+3.4}_{-2.0}$ & $13.4^{+4.4}_{-2.8}$ & 
$14.0^{+1.6}_{-1.4}$ & $13.6^{+1.6}_{-1.0}$ & $13.4^{+1.8}_{-0.8}$ &
$15.8_{-5.4}^{+6.2}$\\

No prior & 
$13.6^{+5.0}_{-2.4}$ & $12.4_{-2.6}^{+4.8}$ &
$14.6^{+1.8}_{-2.2}$ & $14.2^{+1.6}_{-1.6}$ & $13.6^{+1.6}_{-1.2}$ &
$19.8_{-9.0}^{+8.8}$\\
\end{tabular}
\end{center}
\end{table}

\clearpage

\begin{figure}
\epsfxsize=5in
\begin{center}
\leavevmode
\epsffile{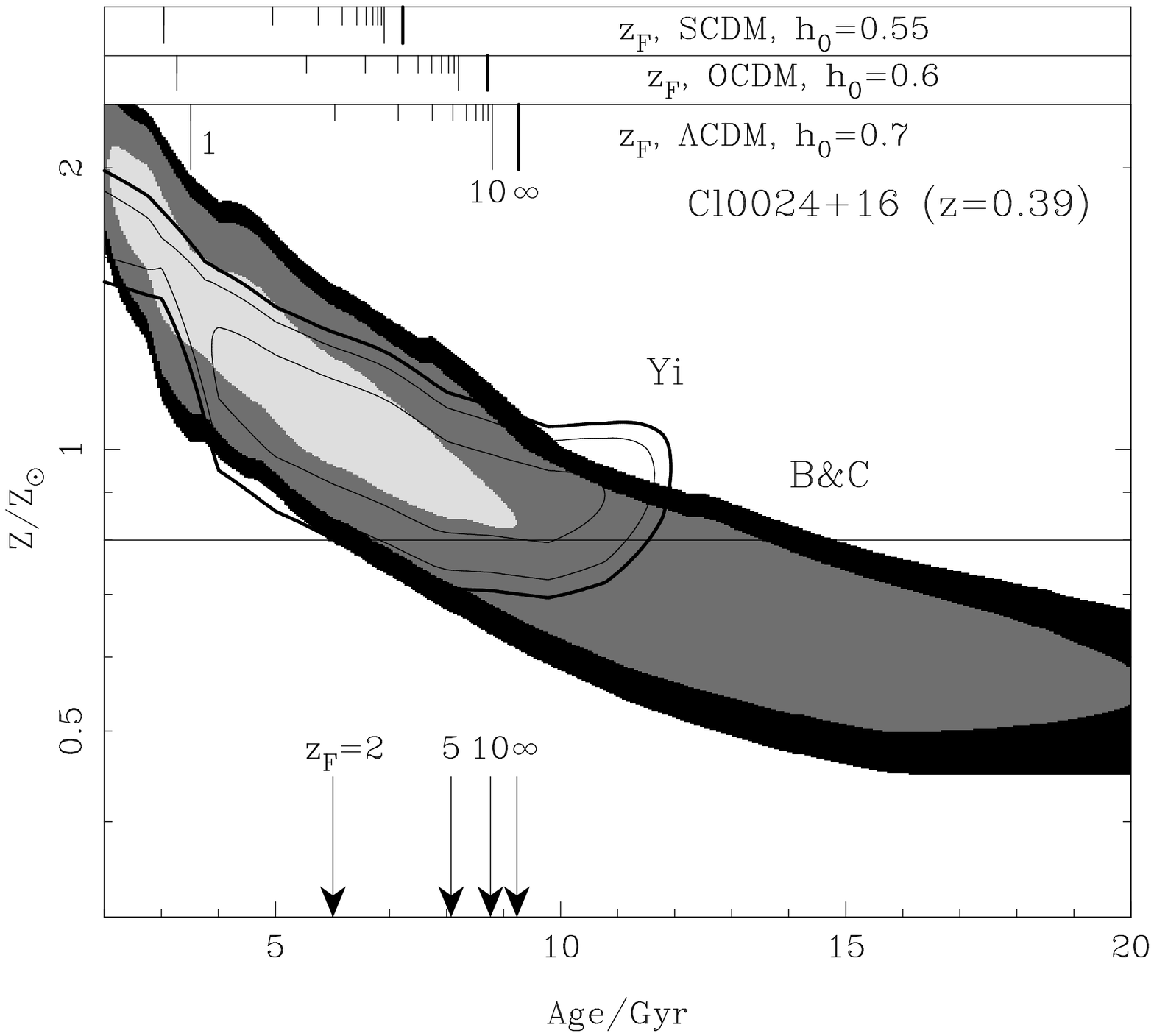}
\bigskip

{\bf Figure 1:}
Age-metallicity diagram for
cluster Cl0024+16 ($z=0.39$) using the photometry from  
Stanford, Eisenhardt \& Dickinson (1998). The contours are at
the 68, 90 and 95\%(thick line) confidence levels of the 
$\chi^2$ defined in the text. Solid (dotted) lines correspond 
to the population synthesis models of Bruzual \& Charlot (2001)
and Yi et al. (1999), respectively. Both assume a Salpeter 
initial mass function. The arrows and one of the top axes give the 
stellar ages for three different formation redshifts assuming 
a $\Lambda$-dominated flat cosmology ($\Lambda$CDM, 
$\Omega_m=0.3$; $h_0=0.7$). The other
axes on top give formation redshifts for other popular
cosmologies: OCDM ($\Omega_m=0.3$; $\Omega_\Lambda =0$; $h_0=0.6$);
and SCDM ($\Omega_m=1$; $\Omega_\Lambda =0$; $h_0=0.55$),
The horizontal line shows the lower
limit to the metallicity expected for bright cluster 
ellipticals.
\end{center}
\end{figure}

\clearpage

\begin{figure}
\epsfxsize=5in
\begin{center}
\leavevmode
\epsffile{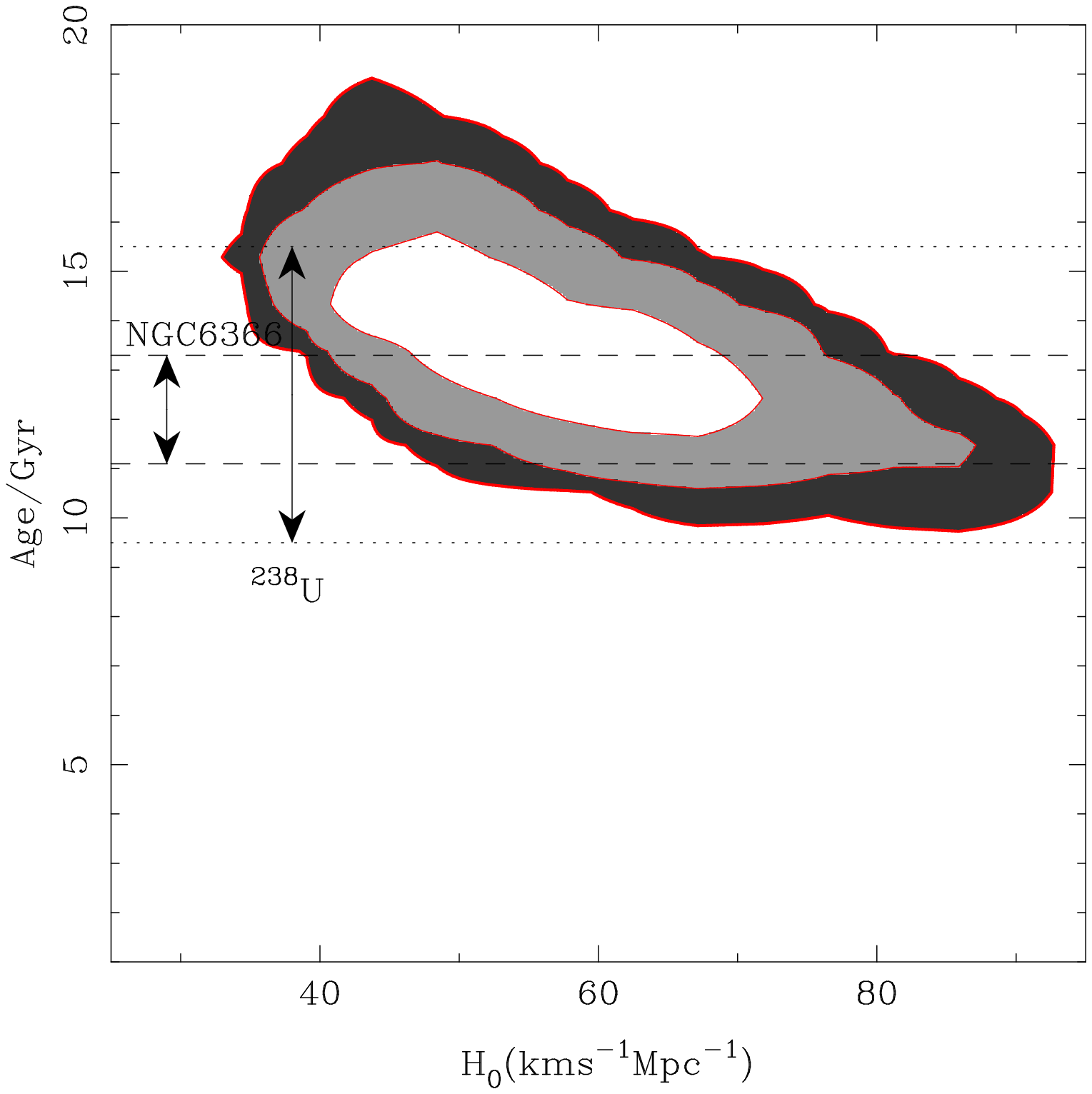}
\bigskip

{\bf Figure 2:}
Estimate of the age of the universe
as a function of $H_0$ preferred by a joint analysis combining
the stellar populations in bright ellipticals for a Salpeter
initial mass function and the observed angular power spectrum of
the CMBR. The $^{238}$U age-measurement
of an old halo star in our galaxy of Cayrel et al. (2001) 
is also shown as dashed lines, however one should shift this
age upwards by an amount corresponding to the lapse between
$t=0$ and the first processes of star formation in our galaxy.
The age of the oldest halo globular cluster in the sample of
Salaris \& Weiss (1998) is also shown.
\end{center}
\end{figure}

\clearpage

\begin{figure}
\epsfxsize=5in
\begin{center}
\leavevmode
\epsffile{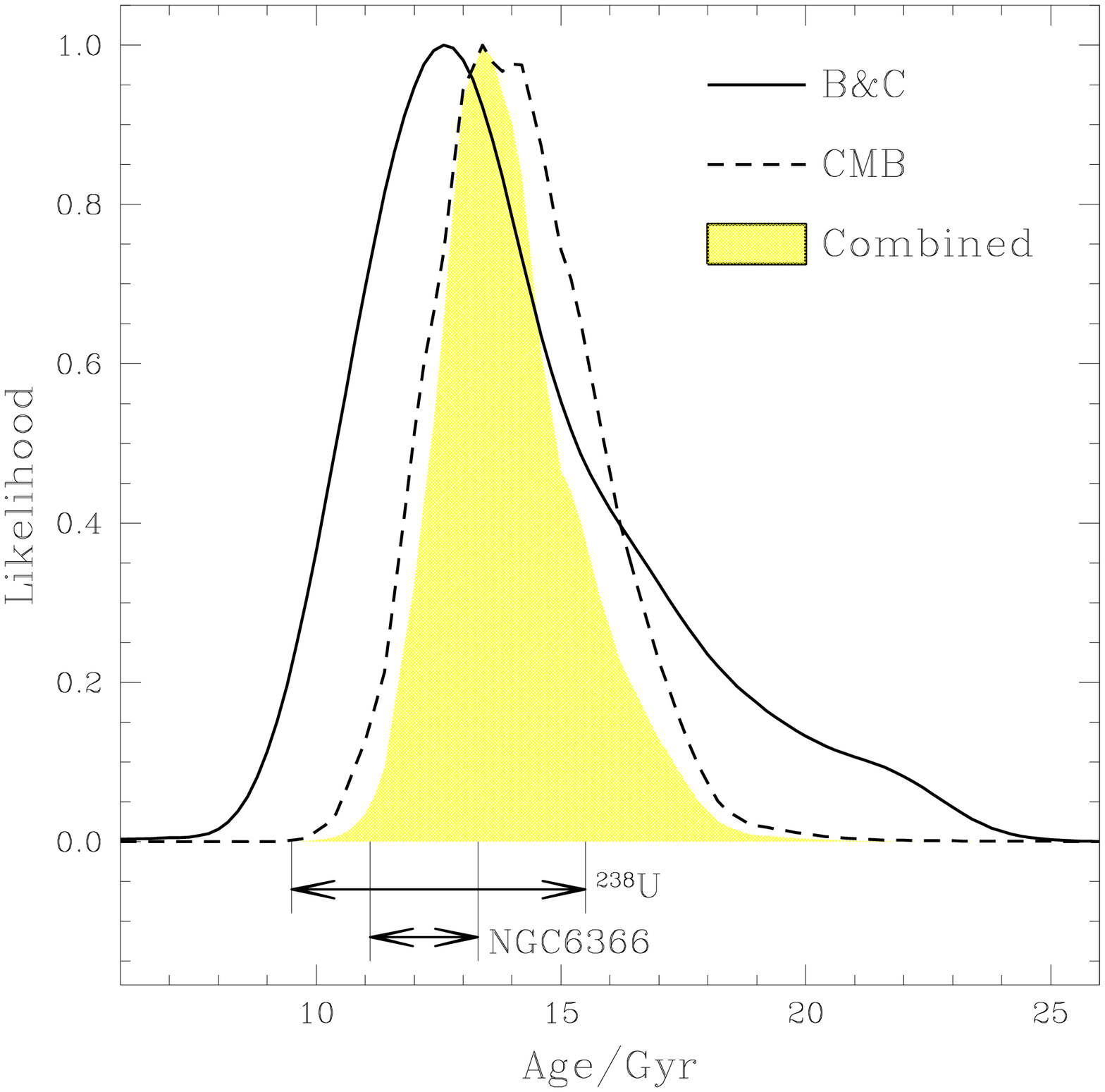}
\bigskip

{\bf Figure 3:}
Likelihood curves for the estimates of the age of the
universe using a comparison between the colours of bright cluster
ellipticals and the latest population synthesis models of 
Bruzual \& Charlot (B\&C; solid line); and the angular power
spectrum of the CMBR (dashed). The shaded area gives the
likelihood of the combined analysis. Universes older than 
$18$~Gyr are ruled out at a 90\% confidence level.
\end{center}
\end{figure}

\end{document}